\documentclass[superscriptaddress,twocolumn,nofootinbib,aps,prl,showpacs,preprintnumbers,reprint]{revtex4-1}
\usepackage{amsmath,amssymb}
\usepackage{epsfig,graphics,color,calc,graphicx}
\usepackage{epstopdf}
\usepackage{hyperref}
\usepackage{mathrsfs}

\usepackage{subfigure}
\usepackage{caption}
\usepackage{color}

\usepackage{ulem}
\begin{document}

\title{Implication of CMS analysis of photon-photon interactions on photon PDFs}

\author{Pazilet Obul}
\email{pazilet.obul@hotmail.com}
\affiliation{
School of Physics Science and Technology, Xinjiang University,
 Urumqi, Xinjiang 830046 China }

\affiliation{
Center for Theoretical Physics, Xinjiang University,
Urumqi, Xinjiang 830046 China}
\author{Mamut Ababekri}
\email{mamutjan@126.com}
\affiliation{
School of Physics Science and Technology, Xinjiang University,
 Urumqi, Xinjiang 830046 China }

\affiliation{
Center for Theoretical Physics, Xinjiang University,
Urumqi, Xinjiang 830046 China}

\author{Sayipjamal Dulat}
\email{sdulat@msu.edu}
\affiliation{
School of Physics Science and Technology, Xinjiang University,
 Urumqi, Xinjiang 830046 China }

\affiliation{
Center for Theoretical Physics, Xinjiang University,
Urumqi, Xinjiang 830046 China}

\affiliation{
Department of Physics and Astronomy, Michigan State University,
 East Lansing, MI 48824 U.S.A. }

\author{Joshua Isaacson}
\email{isaacson@fnal.gov}
\affiliation{
Fermi National Accelerator Laboratory, Batavia, IL 60510 USA. }

\author{Carl Schmidt}
\email{schmidt@pa.msu.edu}
\affiliation{
Department of Physics and Astronomy, Michigan State University,
 East Lansing, MI 48824 U.S.A. }

\author{ C.--P. Yuan}
\email{yuan@pa.msu.edu}
\affiliation{
Department of Physics and Astronomy, Michigan State University,
 East Lansing, MI 48824 U.S.A. }

\begin{abstract}
As part of a recent analysis of exclusive two-photon production of $W^+W^-$ pairs at the LHC, the CMS experiment used di-lepton data to obtain an ``effective'' photon-photon luminosity.
We show how the CMS analysis on their 8 TeV data, along with some assumptions about the likelihood for events in which the proton breaks up to pass the selection criteria, can be used to significantly constrain the photon parton distribution functions, such as those from the CTEQ, MRST, and NNPDF collaborations. We compare the data with predictions using these photon distributions, as well as the new LUXqed photon distribution. We study the impact of including these data on the NNPDF2.3QED, NNPDF3.0QED and CT14QEDinc fits.
We find that these data place a useful and complementary cross-check on the photon distribution, which is consistent with LUXqed prediction while suggesting
that the NNPDF photon error band should be significantly reduced.
Additionally, we propose a simple model for describing the two-photon production of $W^+W^-$ at the LHC. Using this model, we constrain the amount of inelastic photon that remains after the experimental cuts are applied.
\end{abstract}



\maketitle

With the start of the 13 TeV run of the Large Hadron Collider (LHC),
more precise theory calculations are needed to
correctly interpret the present and upcoming experimental data.
Calculations at the next-to-next-to-leading order (NNLO)
 in Quantum Chromodynamics (QCD) are becoming the standard,
so that the theoretical uncertainty
can be reduced to the same order as the experimental uncertainty.
At this level of precision, the leading-order electroweak correction is also important,
because the square of the coupling of the strong interaction ($\alpha_s$) is of
the same order of magnitude as the electromagnetic coupling ($\alpha$).
 Therefore, it becomes necessary to include electroweak corrections in the calculations.

One particular electroweak correction of interest is that due to photons coming from the
proton in the initial state.  This requires the inclusion of the photon as a parton
inside the proton, with an associated parton distribution function (PDF).
This is necessary both for consistency when electroweak corrections are included and
because the photon-initiated processes can become significant at high energies.
The treatment of the photon PDF in a global analysis was first
performed by the MRST collaboration~\cite{Martin:2004dh}.
Since then, both NNPDF and CTEQ collaborations have introduced photon PDFs~\cite{Ball:2013hta,Schmidt:2015zda},
along with PDF evolution at leading order (LO) in QED and next-to-leading order (NLO) or NNLO in QCD.
The MRST2004QED set contains photon PDFs with a parametrization based on radiation off of
``primordial'' up and down quarks, with the photon radiation cut off at either the current quark masses (MRST0), or the constituent quark masses (MRST1) \cite{Martin:2004dh}.
The NNPDF2.3QED set uses a more general photon
parametrization, which was then constrained by Drell-Yan data at the LHC ~\cite{Ball:2013hta}.  This was recently updated in the NNPDF3.0QED set~\cite{Ball:2014uwa}.
The CT14QED set also uses the radiative ansatz, but for the ``inelastic'' component of the photon PDF only and with the inelastic photon momentum fraction at the initial scale left as a free parameter.
Data on isolated photon production in electron-proton deep inelastic scattering (DIS),
measured by the ZEUS Collaboration~\cite{Chekanov:2009dq},
were used to constrain the inelastic initial photon momentum fraction to be less than 0.14\% at the 90\% confidence level (CL) and less than 0.11\% at the 68\% CL~\cite{Schmidt:2015zda}.
In the same article, the CTEQ-TEA group also presented CT14QEDinc sets, which describe the inclusive photon PDF in the proton, given at the initial scale $Q_0$, as the sum of the (inelastic) CT14QED plus the ``elastic'' photon contribution~\cite{Martin:2014nqa}.  The elastic contribution to the photon PDF, in which the initial proton remains intact, was obtained from the Equivalent Photon Approximation (EPA)~\cite{Budnev:1974de}.
Since CT14QEDinc PDFS were obtained from fitting to ZEUS data,
the photon PDFs are better known for the parton momentum fraction $x$ ranging
from $10^{-4}$ to around 0.4.
Recently, a new determination of the photon PDF, LUXqed, was obtained from the lepton-photon structure functions~\cite{Manohar:2016nzj}.
This approach greatly reduces the uncertainties in the determination of the photon PDFs.
Additionally, the NNPDF group recently adopted the LUXqed approach and introduced a new photon PDF, that applies the LUXqed approach to a global PDF fit~\cite{Bertone:2017bme}. Since it yields a result very similar to LUXqed, we will not discuss it further in this work.

With the large amounts of data to be collected at the LHC, photon-initiated processes will become increasingly important.  For instance,
 a precise determination of the quartic couplings of photons and $W$-bosons can be obtained through the analysis of $W$ pair production through
 photon-fusion.  This has been shown to be the most precise channel to measure these couplings~\cite{Pierzchala:2008xc,Chapon:2009hh}, with
 the possibility of measurements that are
several orders of magnitude more precise than the limits found at the Tevatron~\cite{Abazov:2013opa} and LEP~\cite{Belanger:1999aw, Heister:2004yd, Abbiendi:2004bf, Abbiendi:2003jh, Abbiendi:1999aa, Achard:2002iz, Achard:2001eg}.
For all of these uses, a good understanding of the initial photon PDF is vital.

In this paper we consider the CMS studies of exclusive two-photon production of $W$ boson pairs~\cite{Khachatryan:2016mud},
and show how the di-lepton cross-check analysis can be used to constrain the photon PDF.
We compare predictions from the various photon PDFs against each other and against
the CMS data analysis, after invoking a simple model to separate the various photon-photon initiated scattering contributions.
We find that the predictions from various PDF sets are in good agreement with the CMS data under the assumption
that the double dissociative contribution is negligible.
After comparing the photon PDFs of CT14QEDinc, LUXqed, MRST2004QED,
 NNPDF2.3QED, NNPDF3.0QED and NNPDF3.1LUXqed through the photon-photon
luminosity at the LHC with a 13 TeV center-of-mass collider energy,
we demonstrate how the result of the CMS data analysis strongly
constrains the earlier NNPDF2.3QED and NNPDF3.0QED photon PDFs.
Consequently, many studies in the literature that used the NNPDF2.3QED photon
PDF, which predicted large photon-initiated contributions at the LHC (and with large uncertainties
due to the photon PDFs), should see reduced photon-initiated contributions.
As an example, we show that the
predicted high-mass Drell-Yan pair production cross sections at the LHC
are reduced by more than one order of magnitude in the TeV region when the
NNPDF photon PDFs are reweighted to include the impact of the CMS data.

Recently, the CMS experiment at the LHC
has performed measurements of the $W$-boson pair production process ($p p \rightarrow p^{( \ast)} W^+ W^- p^{( \ast)}$) at $\sqrt{s}=7$ TeV~\cite{Chatrchyan:2013akv} and at $\sqrt{s}=8$ TeV~\cite{Khachatryan:2016mud},
and used these to put constraints on anomalous quartic gauge couplings.
In these measurements they selected photon-photon fusion events, including both elastic events,  where both protons remained intact, and inelastic
(quasi-exclusive or ``proton dissociative'') events, in which one or both protons dissociate.  This selection was attained by requiring no additional associated charged tracks
beyond the muon ($\mu$) and electron ($e$) with opposite sign charges
($\mu^\pm e^\mp$), which identified the $W$ boson
pairs, in the central rapidity region (with $|y_{WW}| < 2.5$).
In order to predict the expected rate of $p p \rightarrow p^{( \ast)} W^+ W^- p^{( \ast)}$,
they used the much-higher-statistics sample of $\ell^+\ell^-$ events (away from the $Z$-peak and in the same invariant mass range, with $\ell=\mu$ or $e$) to extract an effective photon-photon luminosity.
This was obtained by taking the ratio of the observed $\ell^+\ell^-$ events with no additional associated charge tracks to that predicted from purely elastic scattering (after subtracting possible quark-initiated contamination,
estimated from $Z$-peak events).  The effective photon-photon luminosity determined from this
data-driven approach was then used to
predict the total cross section to be $\sigma ( p p\rightarrow p^{( \ast)} W^+ W^- p^{( \ast)} \rightarrow p^{( \ast)} \mu^{\pm} e^{\mp} p^{( \ast)}) = 4.0 \pm 0.7$ fb
at $\sqrt{s}=7$ TeV and $6.2 \pm 0.5$ fb at $\sqrt{s}=8$ TeV,
after including $W$ boson decay branching fraction.

\begin{figure}
\centering
\subfigure[Q=3.2 GeV]{
\label{fig-CompQ32} %% label for first subfigure
\includegraphics[scale=0.5]{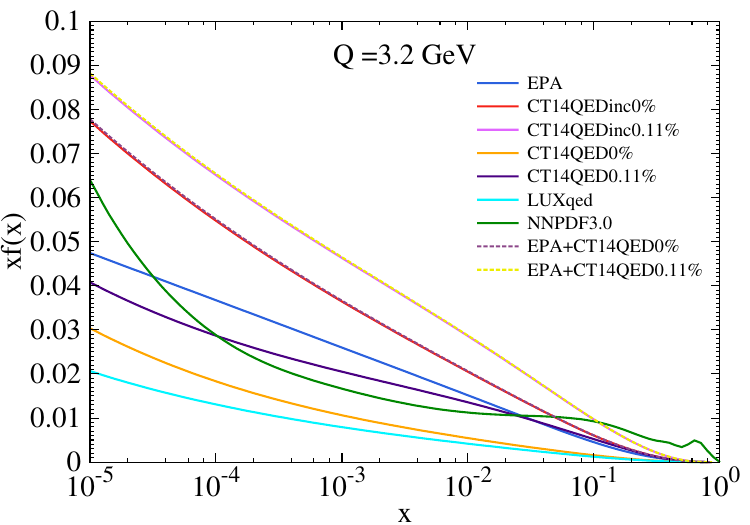}}
\subfigure[Q=100 GeV]{
\label{fig-CompQ100} %% label for second subfigure
\includegraphics[scale=0.5]{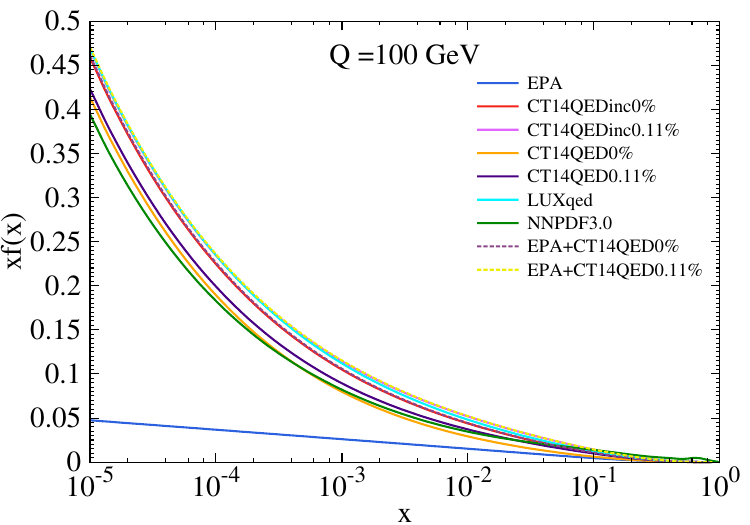}}
\caption{Various elastic (EPA), inelastic (CT14QED) and
inclusive (CT14QEDinc,LUXqed,NNPDF3.0) photon PDF distributions at (a) Q=3.2 GeV and (b) Q=100 GeV.
}
\label{figuer-2-3} %% label for entire figure
\end{figure}

\begin{figure}
\includegraphics[scale=0.5]{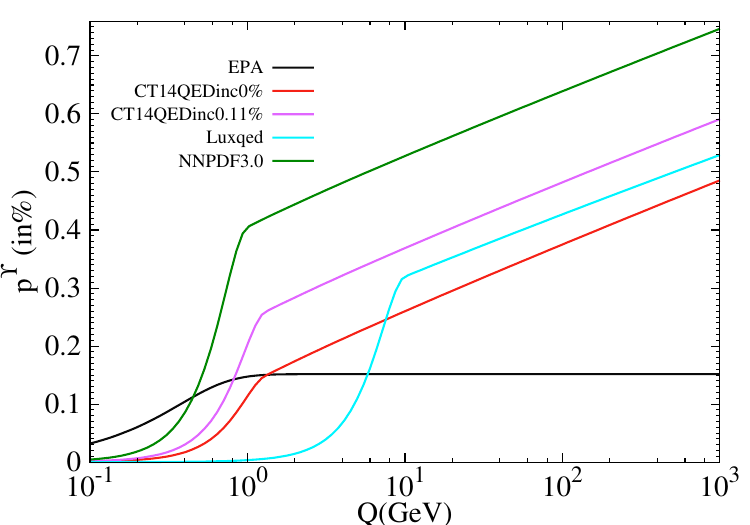}
	\caption{The photon momentum fraction inside the proton as a function of $Q$ for various photon PDFs. The change in
	slope of all the PDFs at the specific low Q value is due to the fact that the PDFs are not defined below some $Q_0$. Below that scale,
	extrapolation is used.}
\label{fig-photonfraction}
\end{figure}

Since these predicted cross sections use their respective extracted photon-photon luminosities, they include both
elastic and  inelastic contributions.  Therefore, they can be used to constrain the photon PDFs if we make some assumptions about the fraction of
dissociative events that pass the no-additional-charged-tracks cut.
For this comparison, we calculate the total cross section for $W$-pair production\footnote{We emphasize that, although we are
using the $W^+W^-$ cross section for the comparison, it is in fact the effective photon-photon luminosity extracted from the CMS di-muon
data that constrains the photon PDFs.} via the photon-photon fusion process
$\gamma \gamma \to W^+ W^-$, with the proper $W$ boson decay branching ratios included,
at the leading-order in electroweak interaction. The factorization scale is
chosen to be the invariant mass ($\sqrt{\hat s}$) of the $W$-boson pair, unless specified.
Using CT14QEDinc PDFs for the inclusive photon
and the EPA for the elastic photon, we separated the prediction into elastic, single-dissociative, and double-dissociative events.
To take into account the cut on additional charged tracks, we use a crude approximation
based on the finding in Ref.~\cite{Harland-Lang:2016apc} that
the probability of not producing extra tracks in the central detector
due to hadronic rescattering is predicted to be
relatively close to 1 for the elastic and single-dissociative cases.
Hence, we assume that the elastic and single-dissociative events all pass the cut, while the double-dissociative events are reduced by a factor $f$, which we
vary between 0 and 1.
Namely, we compare to the effective photon-photon luminosity
extracted from the CMS di-muon data by the following theory
calculation:
\begin{eqnarray} \label{1}
\sigma_{inclusive} & = & \sigma_{elastic} + \sigma_{single-dissociative} \nonumber\\
 & & + f\times \sigma_{double-dissociative}.
\end{eqnarray}
Here, $\sigma_{elastic}$ is calculated using EPA photon PDFs
from both colliding protons; $\sigma_{single-dissociative}$ is obtained by
using one EPA photon PDF and one inelastic photon PDF;
while $\sigma_{double-dissociative}$ is calculated using inelastic photon PDFs from both colliding protons.
The inelastic photon PDF is taken as the difference between an inclusive photon PDF (such as CT14QEDinc, NNPDF3.0QED and LUXqed photon PDFs) and the EPA
photon PDF.
We note that CT14QEDinc PDF includes both elastic and inelastic contributions to the photon PDF,
and can be well-approximated by the linear sum of the elastic component from EPA
and the inelastic component from CT14QED at any given scale $Q$, as illustrated in
Figs.~\ref{fig-CompQ32} and ~\ref{fig-CompQ100}. This observation was used in the original analysis to constrain the
CT14QED and CT14QEDinc photon PDFs~\cite{Schmidt:2015zda} from the ZEUS data, and it also agrees with the
conclusion made in Ref.~\cite{Martin:2014nqa}.
Furthermore, Fig.~\ref{fig-photonfraction} shows that the EPA photon contribution to the proton momentum
($p^\gamma$) becomes essentially constant at scales $Q$ above the initial scale of $Q_0=1.3$ GeV.
The EPA photon PDF is the black curve, while the two CT14QEDinc photon PDFs start at the scale $Q_0=1.3$ GeV
with either 0\% or 0.11\% inelastic photon momentum fraction. For example, at $Q=10$ GeV, the (elastic) EPA photon contributes about 0.15\% of the proton momentum, and
the (inelastic) CTEQ14QED photon contributes about 0.11\% and 0.22\%
of the proton momentum, respectively,
for the two PDF sets labelled by their initial inelastic photon momentum fractions as
[CT14QED 0\%] and [CT14QED 0.11\%].
Hence, at $Q=10$ GeV, the photon momentum fraction of the two corresponding
CT14QEDinc PDFs is about 0.26\% and 0.37\%, respectively.
At 1 TeV, the photon momentum fraction of the NNPDF3.0QED and LUXqed is
 about 0.75\% and 0.53\%, respectively, while
the two corresponding
CT14QEDinc PDFs increase to about 0.48\% and 0.59\%, respectively.

\begin{figure}
\includegraphics[scale=0.35]{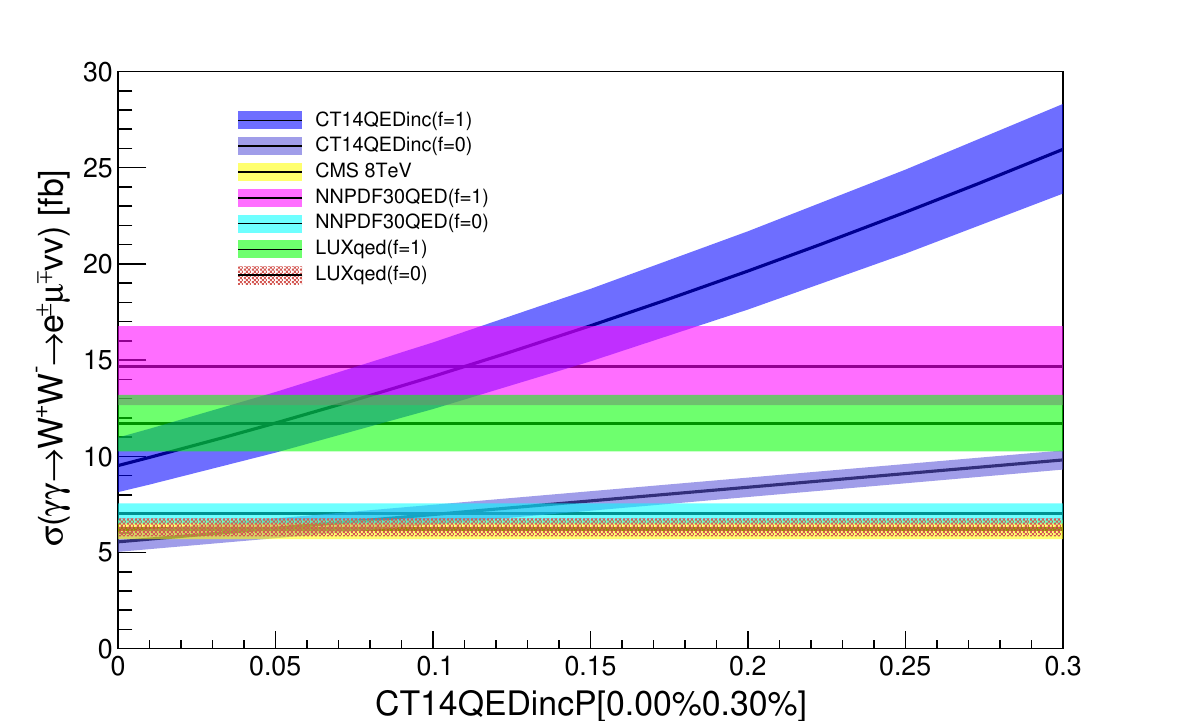}
	\caption{ CT14QEDinc predictions with initial inelastic photon
	momentum fraction varying from 0\% to 0.3\% compared with the CMS result at  $\sqrt{s}=8$ TeV, including uncertainty. Theory bands
	correspond to scale uncertainties between
$\mu=0.5\sqrt{\hat s}$ and $\mu=2\sqrt{\hat s}$.}\label{fig-CT14QED-Xsec8}
\end{figure}

Using this approximation we can calculate the predicted cross section as a function of $f$ and compare with the CMS result.
In Fig.~\ref{fig-CT14QED-Xsec8} we show the predicted cross sections for $f=0$ and $f=1$ using the CT14QEDinc, NNPDF3.0QED and LUXqed PDFs as a function of the initial inelastic photon momentum fraction ($p_0^{\gamma}$) compared with the $\sqrt{s}=8$ TeV prediction from the CMS analysis.
It clearly shows that the
CMS result is consistent with a fraction $f$ much less than 1.
Assuming $f\approx0$, the 8 TeV CMS prediction favors small values of $p_0^{\gamma}\approx0.04\%$ with $p_0^{\gamma}\le0.11\%$ for CT14QEDinc, at the 68\% confidence level (CL). When modelling the cross-section as in Eq.~(1) and assuming $f\approx 0$, the data agree well with predictions based on the LUXqed PDF calculation.
For comparison, we note that this CT14QEDinc result is consistent with the constraint of $p_0^{\gamma}\le0.14$\% at the 90\% CL, derived from comparing to the
 isolated photon production rate in DIS process, measured
 by ZEUS Collaboration~\cite{Schmidt:2015zda}.

\begin{figure}
\includegraphics[scale=0.35]{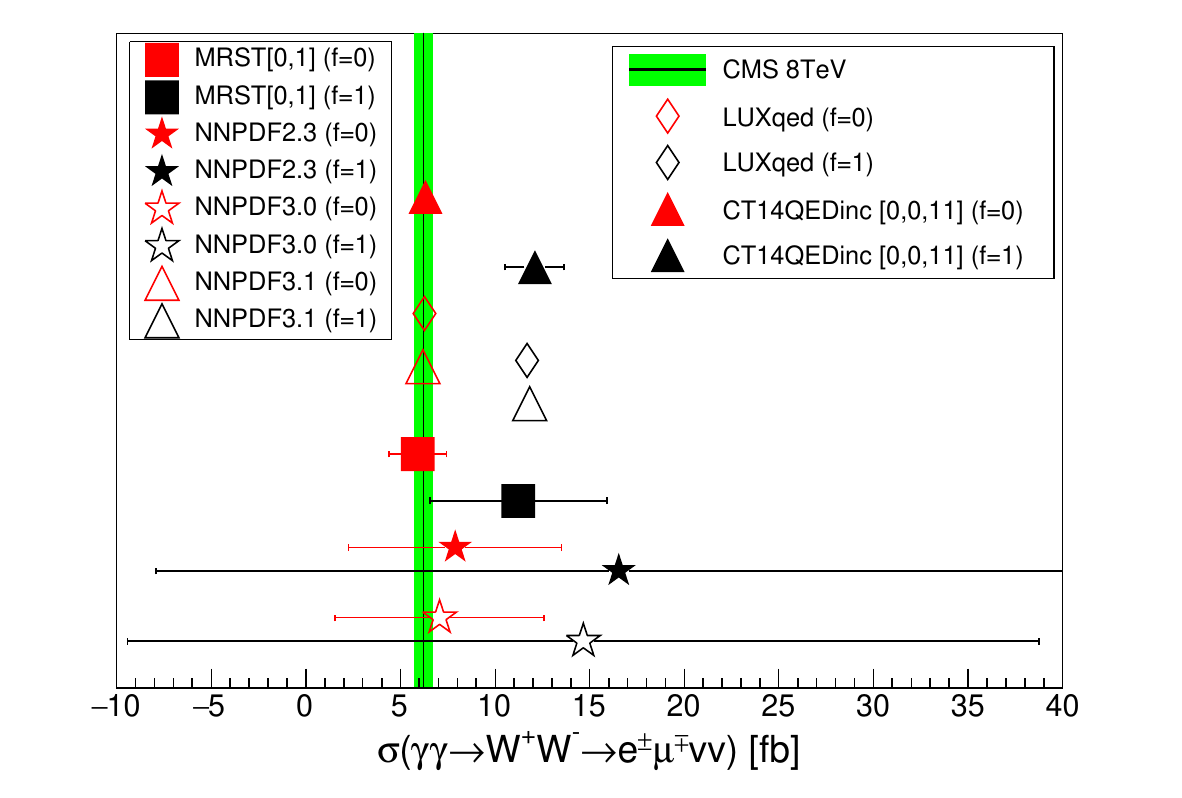}
	\caption{Various PDF set predictions (with their PDF uncertainty ranges) compared to the
  CMS result at 8 TeV, at the 68\% CL.}\label{fig-scaleU8}
\end{figure}

 We can also calculate the same cross section using the other photon PDFs (assumed to be inclusive) in the same manner, as a function of $f$. In Fig.~\ref{fig-scaleU8} we compare the CMS result with predictions from the CT14QEDinc, LUXqed, MRST2004qed, NNPDF2.3QED, NNPDF3.0QED and NNPDF3.1LUXqed photon PDF sets.
In all cases, the $f=0$ assumption is in good agreement with the CMS data.
In addition, we can see that, while all PDF sets are consistent with the data for $f=0$,
the uncertainty due to the photon PDF increases as we change from LUXqed to
CT14QEDinc, MRST, and finally to NNPDF, which predicts the largest uncertainty.
This originates from the different methods used to extract the photon PDFs by the different groups.
LUXqed derived their photon PDF from the proton electromagnetic form factors, obtained partly from data and partly from
theory calculations using PDF4LHC15 PDFs;
CT14QED fit to the ZEUS isolated photon production data, in which
photon-initiated process contributes at the leading order;
MRST2004qed modeled the photon PDF without fitting to data, but using two different scale choices to estimate the uncertainty;
while NNPDF2.3QED and NNPDF3.0QED fit to the inclusive Drell-Yan pair data,
whose production rate is dominated by the much larger quark-antiquark
initiated processes. In other words, the NNPDF2.3QED, NNPDF3.0QED photon PDF fits were dominated by the error
in the measurement of the Drell-Yan pair production rate, which explains the
quite large uncertainty in its Monte Carlo replica sets.

\begin{figure}
  \includegraphics[width=0.40\textwidth]{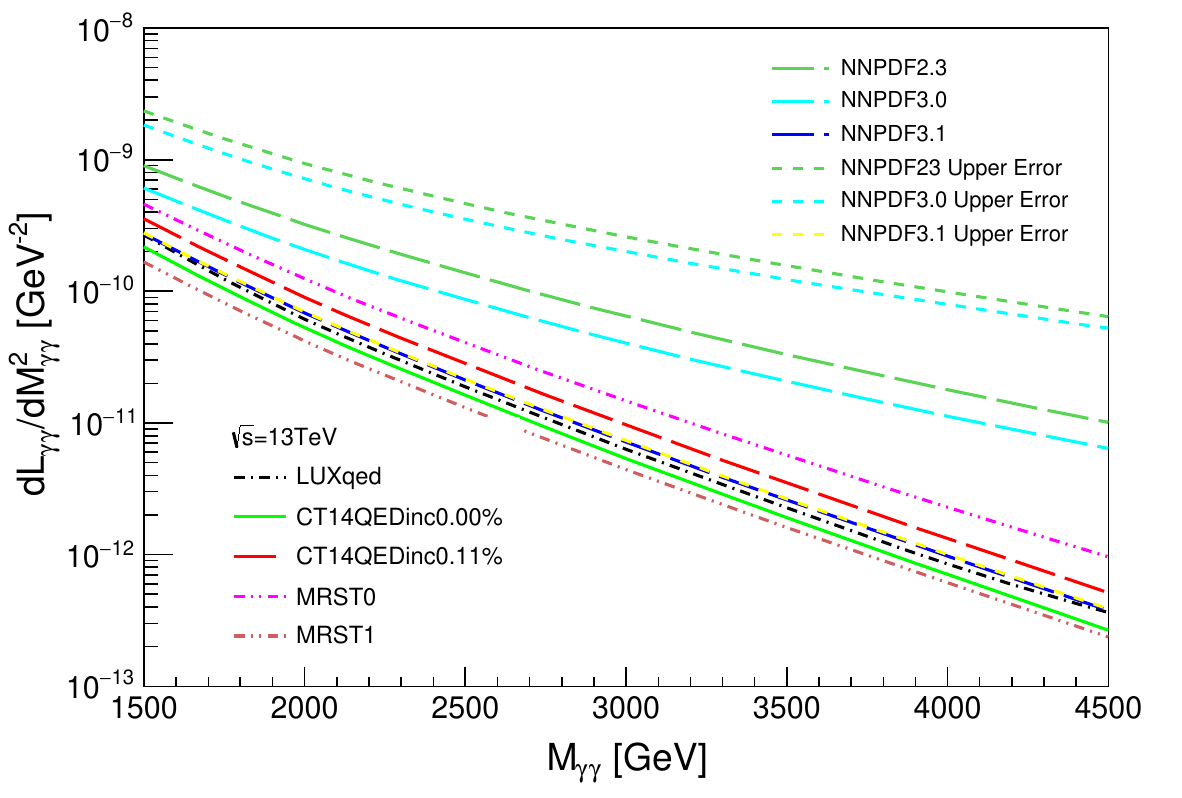}
  \caption{Photon-photon luminosity predicted by various photon PDFs
  for an invariant mass of 1.5 TeV to 4.5 TeV,
  at the LHC with 13 TeV collider energy. The lower error curves of
  NNPDF2.3QED and NNPDF3.0QED
  predictions are below the $x$-axis of this plot.}
  \label{highMass}
\end{figure}

To facilitate the comparison of theory predictions of various production rates induced by the
photon-photon fusion process at the LHC, we compute
the photon-photon parton luminosity for each of the PDF sets, defined as:
\begin{equation}
\frac{d\mathcal{L}_{\gamma \gamma} ( \tau)}{d M^2} = \frac{1}{s}
\overset{}{\underset{}{\int^{- \ln \sqrt{\tau}}_{\ln \sqrt{\tau}}}} dy \,
f_{\gamma / p} ( x_1,\mu_F) f_{\gamma / p} ( x_2,\mu_F)\;,
\end{equation}
where $y=\frac{1}{2} \ln(\frac{x_1}{x_2})$, $\tau = x_1 x_2 = M^2/s$, $M$ is the invariant mass of the photon
pair, and $x_1$, $x_2$ are the momentum fractions of the photons from each proton; the factorization scale $\mu_F$ is chosen to be $M$.
This is shown in Fig.~\ref{highMass} for the LHC at 13 TeV collider energy for the high-invariant mass region. In the high-invariant mass region above approximately 1 TeV, the central NNPDF2.3QED and NNPDF3.0QED luminosities
greatly exceed that of the other PDFs.
This can be traced to the large uncertainty in the photon PDF determination at large $x$,
as well as the extra freedom in the NNPDF2.3QED and NNPDF3.0QED photon PDF parametrization.
Here, we can see that the LUXqed and NNPDF3.1LUXqed luminosity prediction is enveloped by the CT14QEDinc estimated uncertainty,
which in turn is enveloped by the MRST uncertainty, while all of these predictions lie within the NNPDF2.3QED and NNPDF3.0QED error bands.

Next, we examine the impact of the CMS data on the CT14QEDinc, NNPDF2.3QED
and NNPDF3.0QED photon PDFs.
We adopt the PDF Bayesian reweighting technique to study its effect.
The idea of reweighting PDFS was originally proposed by  Giele and Keller in~\cite{Giele:1998gw},
and later
discussed by the NNPDF collaboration~\cite{Ball:2010gb,Ball:2011gg}.
In Ref.~\cite{Sato:2013ika}, a detailed discussion was given to compare  these two reweighting methods and
favored the original procedure in~\cite{Giele:1998gw}.
(In the case of including only one new data point, such as in the present study, both methods
coincide.)
 The reweighting technique assigns weights to each of the replica sets, which strongly suppress those
whose theory predictions are in poor agreement with the new (CMS) data.
The weights are derived from the chi-square ($\chi^2$) values of the comparison between the new data and
theory prediction from each of the PDF replicas.
The central value of any observable is the weighted
average of the values extracted from each of the
PDF replicas, and its PDF error is given by the weighted root-mean-square (RMS) of those values~\cite{Giele:1998gw}.
While NNPDF2.3QED and NNPDF3.0QED are already in the form of Monte Carlo replicas, we need to first
construct the Monte Carlo replicas from the two CT14QEDinc photon PDFs,
 [CT14QEDinc 0\%] and [CT14QEDinc 0.11\%], which represent the two error PDFs
along the negative and positive direction of the photon error PDF eigenvector
in Hessian method~\cite{Pumplin:2001ct}.
For that, we use the public code MCGEN~\cite{MCGENwebsite} which facilitates the method described in Ref.~\cite{Hou:2016sho} to
generate the CT14QEDinc replicas for this study.

\begin{figure}
\centering
\subfigure[NNPDF-PR]{
\label{fig-NNPDFreweight} %% label for first subfigure
\includegraphics[scale=0.35]{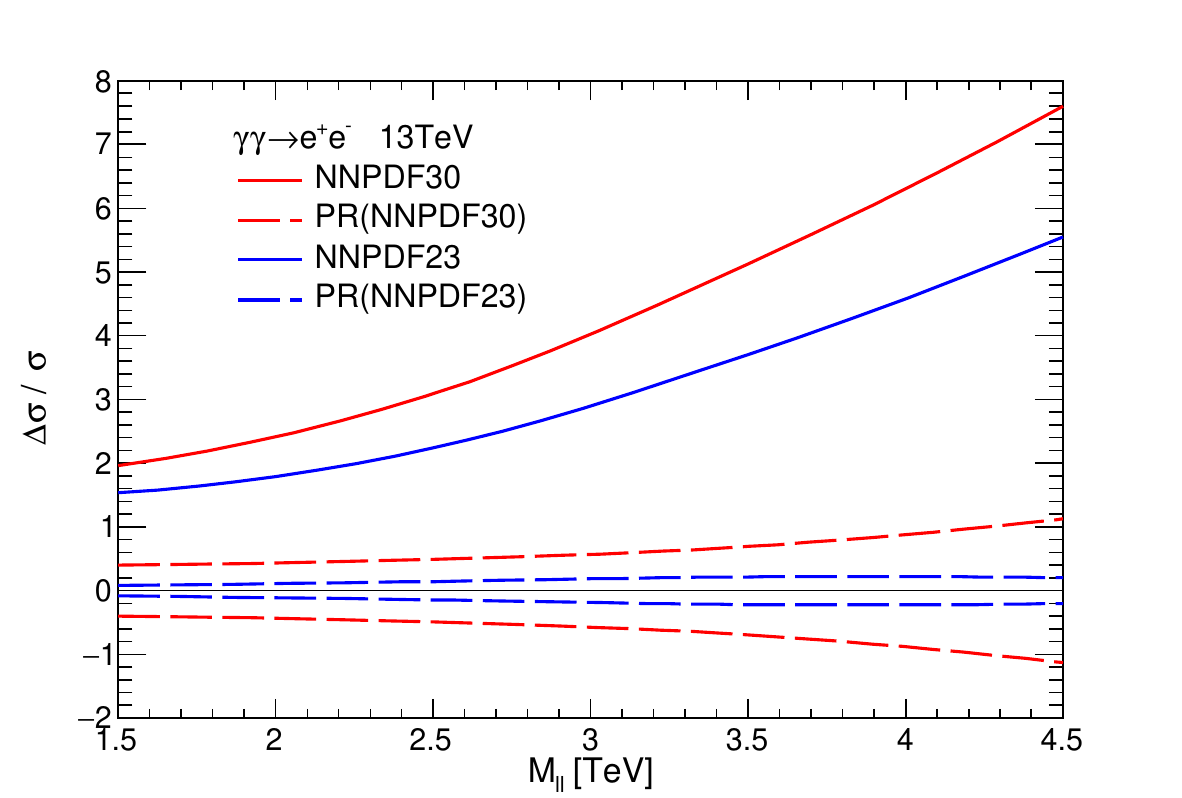}}
\subfigure[CT14-PR]{
\label{fig-CT14reweight} %% label for second subfigure
\includegraphics[scale=0.35]{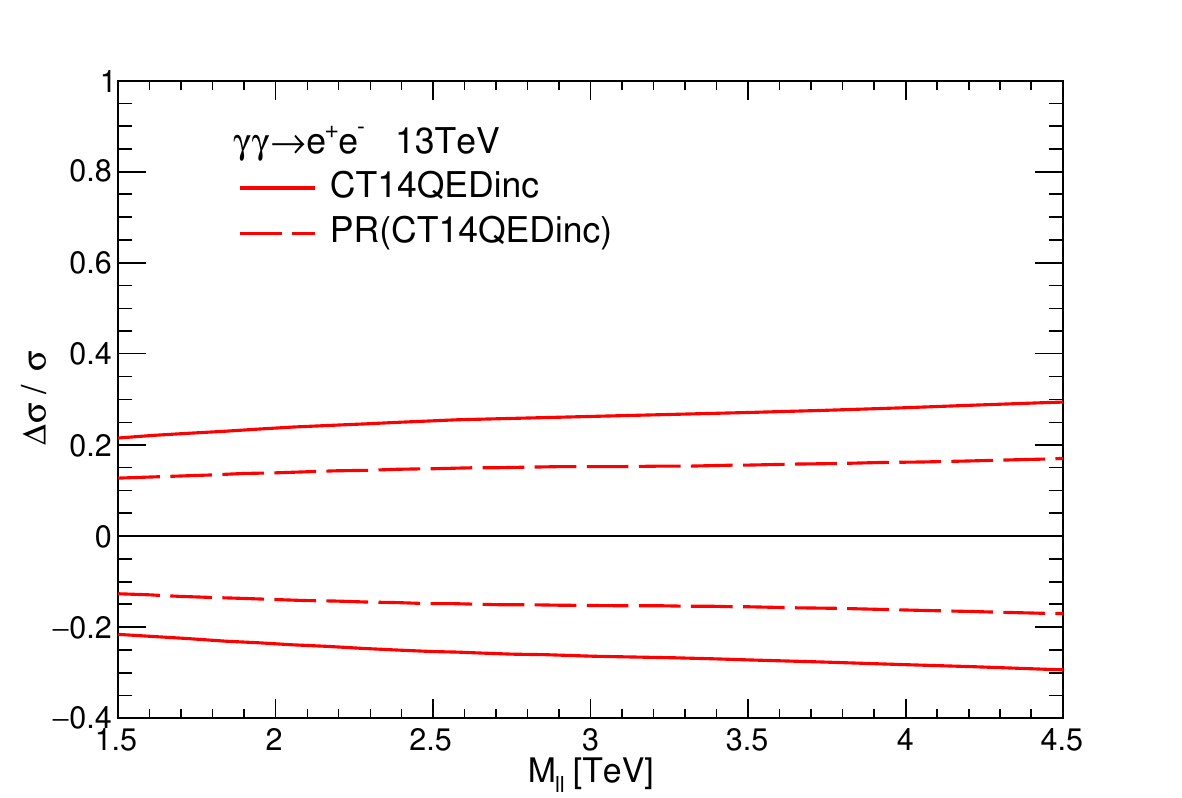}}
\caption{The (a) NNPDF2.3QED and NNPDF3.0QED, and (b)
CT14QEDinc photon PDF induced uncertainties in the lepton pair invariant mass distribution,
via $\gamma \gamma \to e^-e^+$ at the 13 TeV LHC,
before and after PDF-reweighting (PR).}
\label{figuer-NN-CT-PR} %% label for entire figure
\end{figure}

\begin{figure}
	\includegraphics[width=0.40\textwidth]{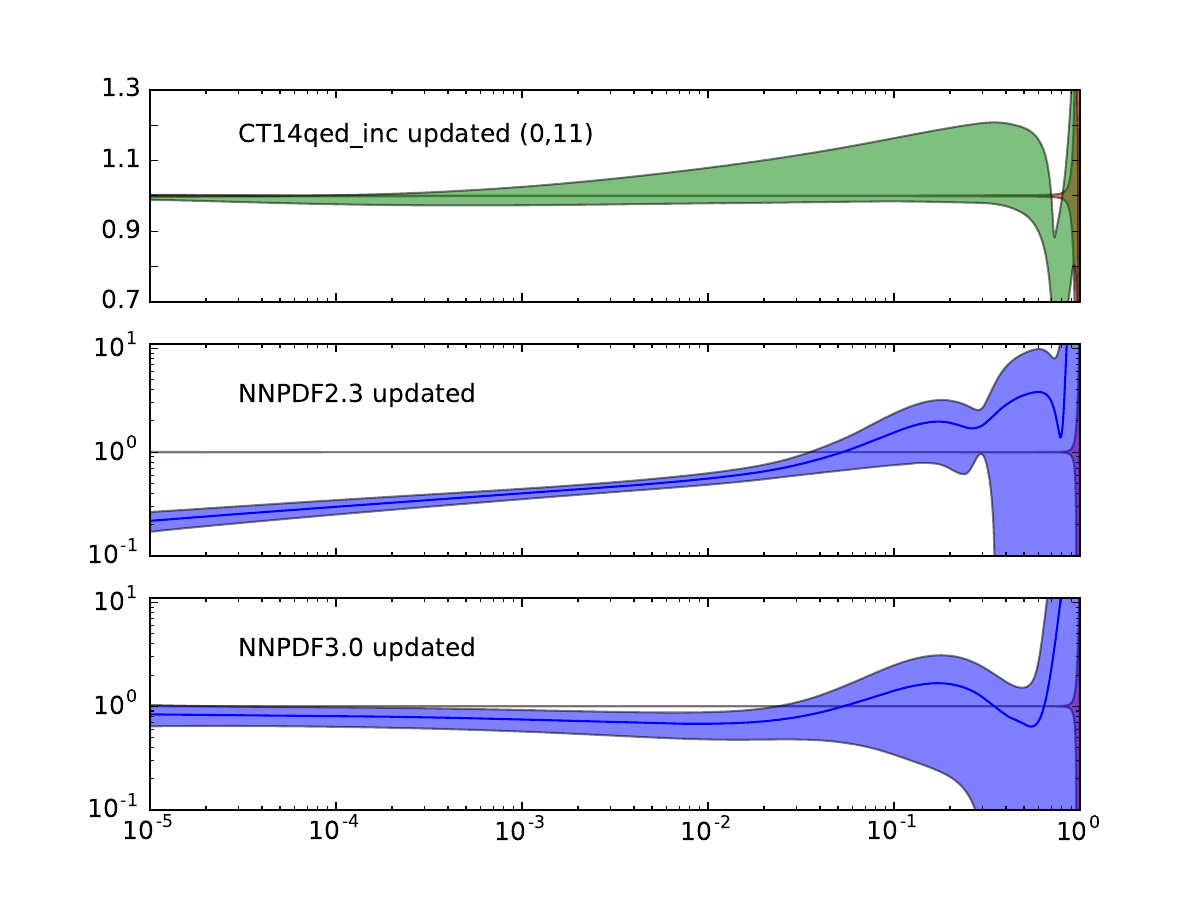}
	\caption{The ratio of common PDF sets to LUXqed result, along with the LUXqed uncertainty band (light red),
		after imposing the constraint from the CMS data at the 68\% cl.
		The CT14 band corresponds to the range from the PDF members shown in brackets after reweighting. The NNPDF bands are calculated using the reweighted replicas. The uncertainty is given by the standard deviation of the updated replicas.
		Note the different $y$-axes for the panels.}
	\label{comp-pdfs}
\end{figure}

\begin{figure}
  \includegraphics[width=0.40\textwidth]{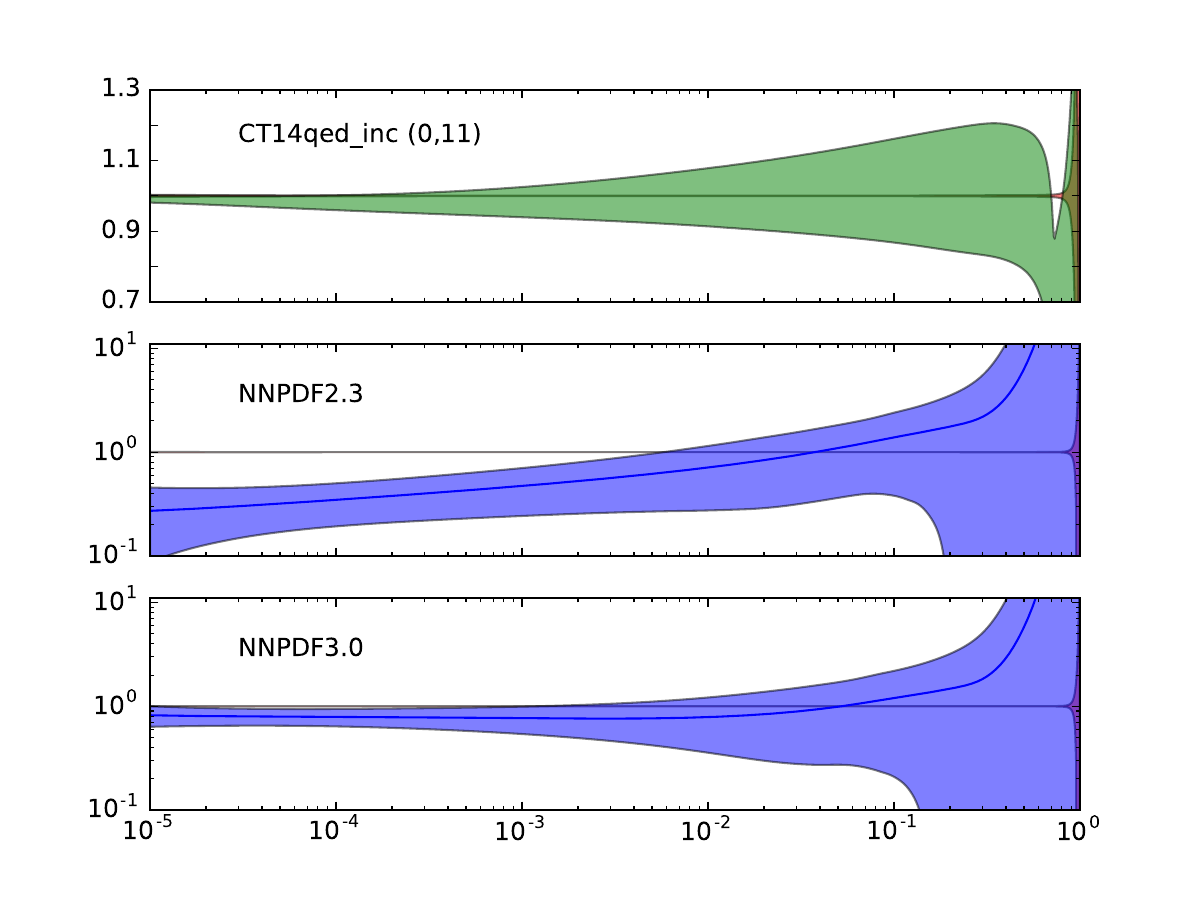}
  \caption{The ratio of common PDF sets to LUXqed result, along with the LUXqed uncertainty band (light red),
  before imposing the constraint from the CMS data at the 68\% cl.
  The CT14 band corresponds to the range from the PDF members shown in brackets after reweighting. The NNPDF bands are calculated using the reweighted replicas. The uncertainty is given by the standard deviation of the updated replicas.
  Note the different $y$-axes for the panels.}
  \label{comp-pdfs-preupdate}
\end{figure}

The results of including the CMS data to reweight the different photon PDF replicas are shown in Figs.~\ref{fig-NNPDFreweight} and \ref{fig-CT14reweight},
where we calculate the relative uncertainties in the distribution of lepton pair invariant mass in the high mass region.
As expected, the PDF uncertainties for this distribution are reduced for both the CT14QEDinc and NNPDF photon PDF sets after including the 8 TeV CMS data.
In particular, the CMS data can have a very large effect in reducing the errors due to the NNPDF photon PDFs.
For example, at 2 TeV and 3 TeV, the relative errors ($\Delta \sigma / \sigma$)
in the NNPDF3.0QED predictions reduce from 240\% and 380\%, respectively,
to about 40\%, while the average values of the cross sections ($\sigma$) reduce by about a
factor of 2 after including the 8 TeV CMS data.
In contrast, the reduction in $\Delta \sigma / \sigma$ in
the CT14QEDinc prediction is mild, from about 25\% to 15\%,
while the average predicted $\sigma$ is almost unchanged.
For completeness, we also show in Fig.~\ref{comp-pdfs} the comparison of various photon PDFs, similar to Fig. 4 of Ref.~\cite{Bertone:2017bme}, but after imposing the constraint from the CMS data. 
We note that for the NNPDF sets we always use the standard deviation for the uncertainty instead of taking the $\textrm{max}\left(\mu-\sigma,r_{16}\right)$, where $r_{16}$ is the replica at the $16^{\textrm{th}}$ percentile, as done in Fig. 4 of Ref.~\cite{Bertone:2017bme}.   
For comparison, in Fig.~\ref{comp-pdfs-preupdate}, we show the PDFs before they are updated by the CMS data.

The CMS data can also be used to test the above proposed model. Based on the cuts used by CMS and the LUXqed PDF set, the
95\% confidence limit for $f$ is given as 0.08. This value can be used as a conservative estimate for the theoretical uncertainty of
the amount of double-dissociative events that pass the no additional track cut.
In a more complete study, where the elastic and single-dissociative events are not assumed to be
fully accepted by the no-extra-track cut, the value of $f$ will be somewhat larger.
We further leave the more detailed analysis to a future work.

In summary, we have shown that the ``effective'' photon-photon luminosity
obtained by the CMS collaboration from analyzing the exclusive two-photon
production of $W^+W^-$ pairs at the LHC can constrain some photon PDFs, particularly, NNPDF2.3QED and NNPDF3.0QED photon PDFs. 
On the other hand, the uncertainty predicted by LUXqed PDFs, with $f=1$, is well within the experimental error of the CMS data.
Many previous analyses that were based on NNPDF2.3QED or NNPDF3.0QED photon PDFs and
had found a large contribution from photon-induced processes need to be reexamined.
For example, it is pointed out in Ref.~\cite{LHCHXSWG} that the largest source of
uncertainty for predicting the $W^\pm H$ production rate, which is important for
measuring the coupling of Higgs boson to $W$ bosons, is due to photon-induced
contributions. This conclusion needs to be reexamined, based on our finding
that the NNPDF photon PDFs overestimate the photon contribution to, as well as the uncertainty in,
 the calculation of processes such as $W^\pm H$,
lepton-pair or vector-boson-pair production at the LHC.
Likewise, it will also modify early conclusions about
the potential of the LHC and future hadron colliders to search for new physics
effects induced by photon-initiated process, {\it e.g.,} Ref.~\cite{Alva:2014gxa}.

\vspace{8mm}
We thank Tao Han, Lucian Harland-Lang, Joey Huston, Valery Khoze,
Wayne Repko, Richard Ruiz and Misha Ryskin for helpful discussions.
We also thank Tie-Jiun Hou and Pavel Nadolsky for providing the
Monte Carlo replicas of CT14QEDinc photon PDFs.
This work was supported by the U.S. National
Science Foundation under Grant No. PHY-1417326 and PHY-1719914;
and by the National Natural Science Foundation of China under Grant
No. 11465018.
C.-P. Yuan is also grateful for the support from
the Wu-Ki Tung endowed chair in particle physics.

\end{document}